\documentclass[%
 aip,
 jmp,%
 amsmath,amssymb,
 reprint,%
 groupedaddress,
 twocolumn
]{revtex4-1}

\usepackage{graphicx}
\usepackage{dcolumn}
\usepackage{bm}

\graphicspath{{figs/}}

\usepackage{caption}
\usepackage[document]{ragged2e}

\usepackage{latexsym}
\usepackage{graphicx, graphics, hyperref, amsmath, amssymb, slashed, xcolor, bbm,amsthm, array,textcomp,gensymb}
 \usepackage{subfigure}
\usepackage{rotating}
\usepackage{afterpage}

\newcommand{\nc}{\newcommand}
\nc{\beq}{\begin{equation}}
\nc{\eeq}{\end{equation}}
\nc{\barray}{\begin{eqnarray}}
\nc{\earray}{\end{eqnarray}}
\nc{\barrayn}{\begin{eqnarray*}}
\nc{\earrayn}{\end{eqnarray*}}
\nc{\bcenter}{\begin{center}}
\nc{\ecenter}{\end{center}}
\nc{\mc}{\mathcal}
\nc{\er}[1]{(\ref{eq:#1})}
\nc{\onehalf}{\frac{1}{2}} 
\nc{\partialbar}{\bar{\partial}}
\nc{\psit}{\widetilde{\psi}}
\nc{\Tr}{\mbox{Tr}}
\nc{\hc}{\mbox{H.c.}}
\nc{\ev}{\;\mathrm{eV}}
\nc{\mev}{\;\mathrm{MeV}}
\nc{\gev}{\;\mathrm{GeV}}
\nc{\tev}{\;\mathrm{TeV}}

\def\chii0{\chi_i^0}
\def\chij0{\chi_j^0}

\newcommand{\gsim}{\lower.7ex\hbox{$\;\stackrel{\textstyle>}{\sim}\;$}}
\newcommand{\lsim}{\lower.7ex\hbox{$\;\stackrel{\textstyle<}{\sim}\;$}}
\nc{\ttbar}{t\bar t}

\newcommand{\cref}[1]{Chapter~\ref{c.#1}}
\newcommand{\tref}[1]{Box~\ref{b.#1}}

\begin{document}


\title{
Flashes of Hidden Worlds at Colliders
}

\author{David Curtin}
\author{Raman Sundrum}

\affiliation{Maryland Center for Fundamental Physics, Department of Physics, University of Maryland, College Park, MD 20742 USA}

\date{\today}

\maketitle

\begin{quotation}

Spectacular bursts of particles appearing seemingly out of nowhere could shed light on some of Nature's most profound mysteries. 
\end{quotation}
\vspace*{3mm}


\justify 

Why is there something rather than nothing?
This is a question so basic, and so vast in scope, that it seems beyond the realm of quantitative enquiry.
And yet, surprisingly, it is a question that can be be framed, and at least partially answered, in the study of particle physics.  

The Standard Model (SM) is our current theory of matter and its interactions, written in the language of relativistic quantum field theory (QFT). It does a great job of explaining the visible world we see around us, and stands up to several decades of tests at high energy colliders. We know, however, that it cannot be the complete description of the universe, in large part because it fails to answer the above question in several important regards. 

The existence of ``something'' is connected to three fundamental mysteries. 
Firstly, why can matter clump to form a rich array of structures in the universe, without collapsing into black holes?
The requisite weakness of gravity derives from the fact that particles which make up everyday matter are \emph{incredibly light}. 
In the SM, the mass of all fundamental particles must be less than the \emph{Electroweak Mass} $\sim$ a few hundred GeV$/c^2$, derived from the physics of the Higgs Boson. 
However, once we take the effects of gravity into account, the Electroweak Scale receives typical quantum mechanical contributions of order the \emph{Planck Mass} $\sqrt{\hbar c/8 \pi G_\mathrm{Newton}} \sim 10^{18} \gev/c^2$.
This puzzling discrepancy, reconciled in the SM only via an incredible cancellation amongst unrelated parameters, is called the \emph{Hierarchy Problem}. 
There are attractive theoretical solutions, such as Supersymmetry or Higgs Compositeness, which predict high-energy signatures of new physics at the Large Hadron Collider (LHC).

Secondly, at some point in the early universe, there was one extra particle of SM matter for every one billion matter-antimatter pairs. This tiny excess is all that remains today, and it must have been created dynamically in the primordial plasma of the Big Bang. In the SM, such a process of \emph{baryogenesis} is too weak by many orders of magnitude, and there must be additional particles and interactions. There are many possibilities for what these might be, including additional Higgs Bosons and modified Higgs couplings which we could detect at the LHC. 

Finally, there's \emph{Dark Matter} (DM), for which there is strong interlocking evidence from a multitude of astrophysical and cosmological observations. 
It makes up about 80\% of the matter in our universe, and yet we have no idea what it is, or how it connects to the SM.
The most popular theories of DM also predict a tiny interaction with ordinary matter, allowing for \emph{direct detection} by nuclear recoils which we can look for in sensitive detectors, as well as \emph{indirect detection} in cosmic ray data of DM annihilation into SM particles.

These three fundamental questions drive much of experimental and theoretical particle physics. 
Many experimental searches for answers are proceeding vigorously, but so far with null results across the board. 
That doesn't mean there's nothing to be discovered. On the contrary, these null results may well point us towards the true nature of the universe. 
There could be hidden sectors -- additional particles and forces -- with only tiny couplings to our SM. 
Far from being inconsequential, these new sectors can address all of the three big mysteries.
Their signatures are subtle and easily missed, but luckily, their hidden nature is also the key to their discovery.
Invisible \emph{Long Lived Particles} (LLPs) can be produced at colliders and decay into energetic SM particles after traveling some distance. 
Can we catch these revealing flashes?

\section*{A tale of two frontiers}

{
We build colliders of ever-increasing energy, like the LHC, to access new physics at higher mass scales than the SM. However, it is also possible for new physics to be concealed not by virtue of high mass, but because its couplings to the SM are very small. We refer to such disconnected non-SM matter and forces, which can robustly arise within the framework of QFT, as \emph{Hidden Sectors}. 
}

Like the push for higher energy, the discovery of hidden sectors with tiny couplings to visible particles has important historical precedent. Neutrinos are almost massless, and nowadays understood to be related to charged leptons (electrons, muons and taus) via their shared weak interaction. {{However, because the force carriers of the electroweak interaction have masses of order $100 \gev/c^2$, the effective coupling of neutrinos to charged leptons is very small at low energies.}} Wolfgang Pauli postulated their existence in 1930 to restore conservation of momentum in beta decay, but their direct detection via scattering off nuclei wasn't achieved until 1956, when large neutrino fluxes from nuclear reactors became available.  The discovery of this ghost-like neutrino sector was only the beginning, and these elusive particles still aren't done confusing us. For example, we know that neutrinos have masses and oscillate from one flavor to another, which the SM does not account for.

\begin{table}
\begin{center}
\fbox{
\begin{minipage}{0.45\textwidth}
\justify
\begin{center}
\includegraphics[width=0.9 \textwidth]{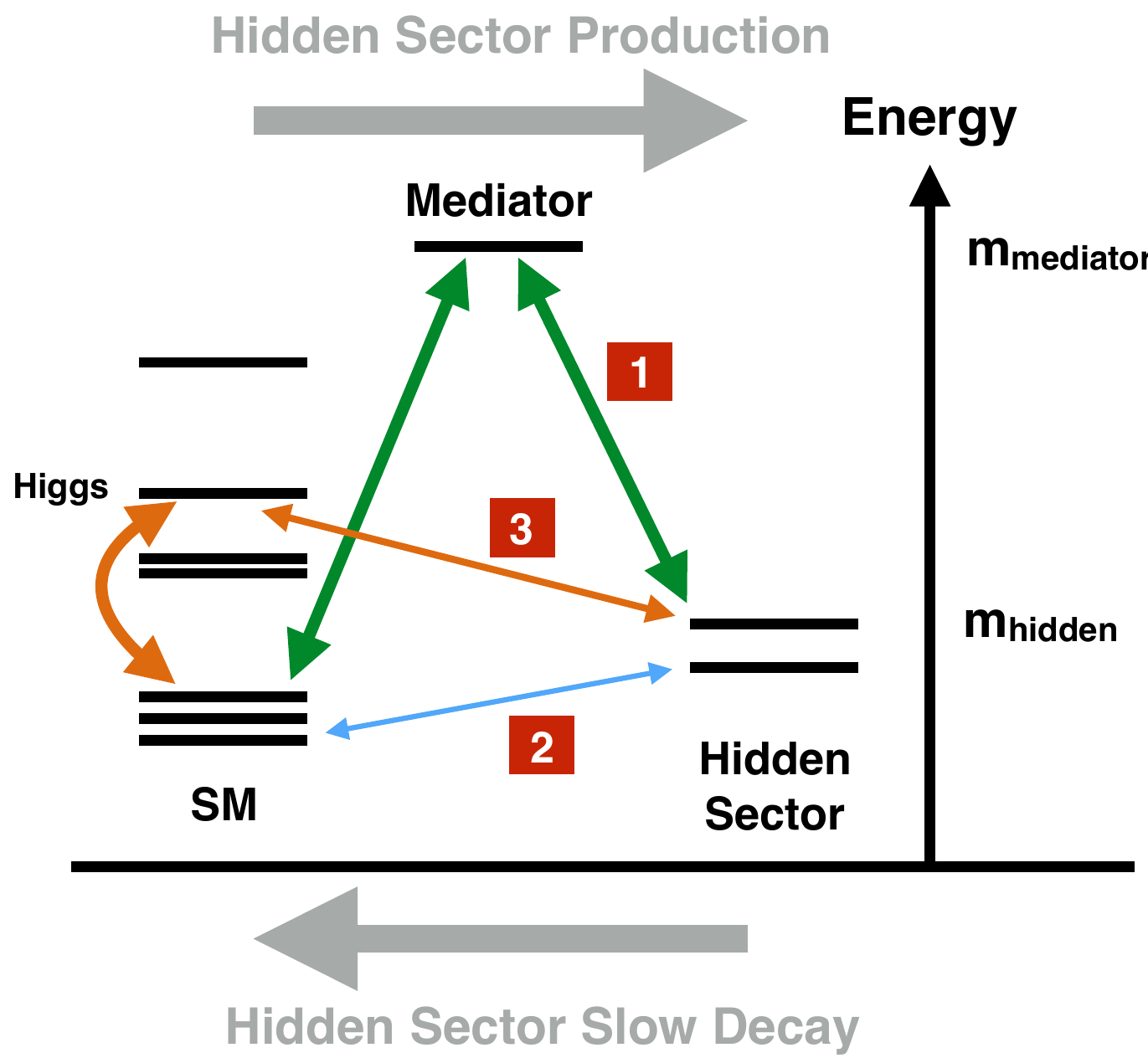}
\end{center}
\normalsize 

{
Schematic representation of the states of a possible Hidden Sector in relation to the SM. Colored arrows indicate possible transitions between states.}

{
Hidden Sector states can be created via the production and decay of heavy mediators [1] at the LHC, via small direct couplings [2], or via exotic decays of the Higgs boson [3]. Once produced, they decay through the same portals. This naturally leads to long lifetimes since the direct couplings are small, or since it requires energy to be ``borrowed'', courtesy of the Heisenberg Uncertainty Principle, to excite the intermediate mediator or Higgs boson. }

\end{minipage}
}
\end{center}
\vspace*{-4mm}
\caption{\textbf{Portals to Hidden Sectors}}
\label{b.atomictransitions}
\end{table}

{New} hidden sectors can be connected to the SM via small but nonzero effective couplings called \emph{portals}. 
This is especially motivated if the hidden sectors play a part in solving some of the big mysteries. 
The most important portal types are illustrated in \tref{atomictransitions}. 
These couplings are small for a variety of different reasons. 
Symmetries can give rise to quantum mechanical \emph{selection rules}, forcing interactions between two sectors to proceed via an intermediate heavy \emph{mediator} state. 
The mediator is not part of the SM but interacts with both sectors.
Symmetries can also be approximate, allowing only small couplings which violate them.

It's  possible for some SM states to play the role of the mediator, most importantly the photon or the Higgs boson. 
While the structure of the theory makes these \emph{Higgs Portal} or \emph{Photon Portal} couplings smaller than ordinary SM couplings, they are readily much larger than other types of portals. 
Furthermore, we already make lots of Higgs bosons and photons!
In rough analogy to neutrino oscillations, the photon could transform into a hidden photon\cite{Bjorken:2009mm} and interact with hidden states, while a Higgs Boson, with a mass of 125 GeV$/c^2$, is heavy enough to decay directly into the Hidden Sector some of the time. Such \emph{exotic Higgs decays} are one of the most promising avenues for producing hidden sector particles.\cite{Strassler:2006ri, Curtin:2013fra}

Hidden sectors typically contain massive states which would be stable in isolation, but in the presence of portal couplings, they are unstable and decay to the SM. 
Precisely because the portal is such a tiny keyhole, this decay can take a relatively long time!
 This is what makes \emph{Long-Lived Particles} and their spectacular decays a hallmark of hidden sectors.

\begin{table*}
\begin{center}
\hspace*{-7mm}
\fbox{
\begin{minipage}{1.06\textwidth}
\begin{center}
\includegraphics[width=1 \textwidth]{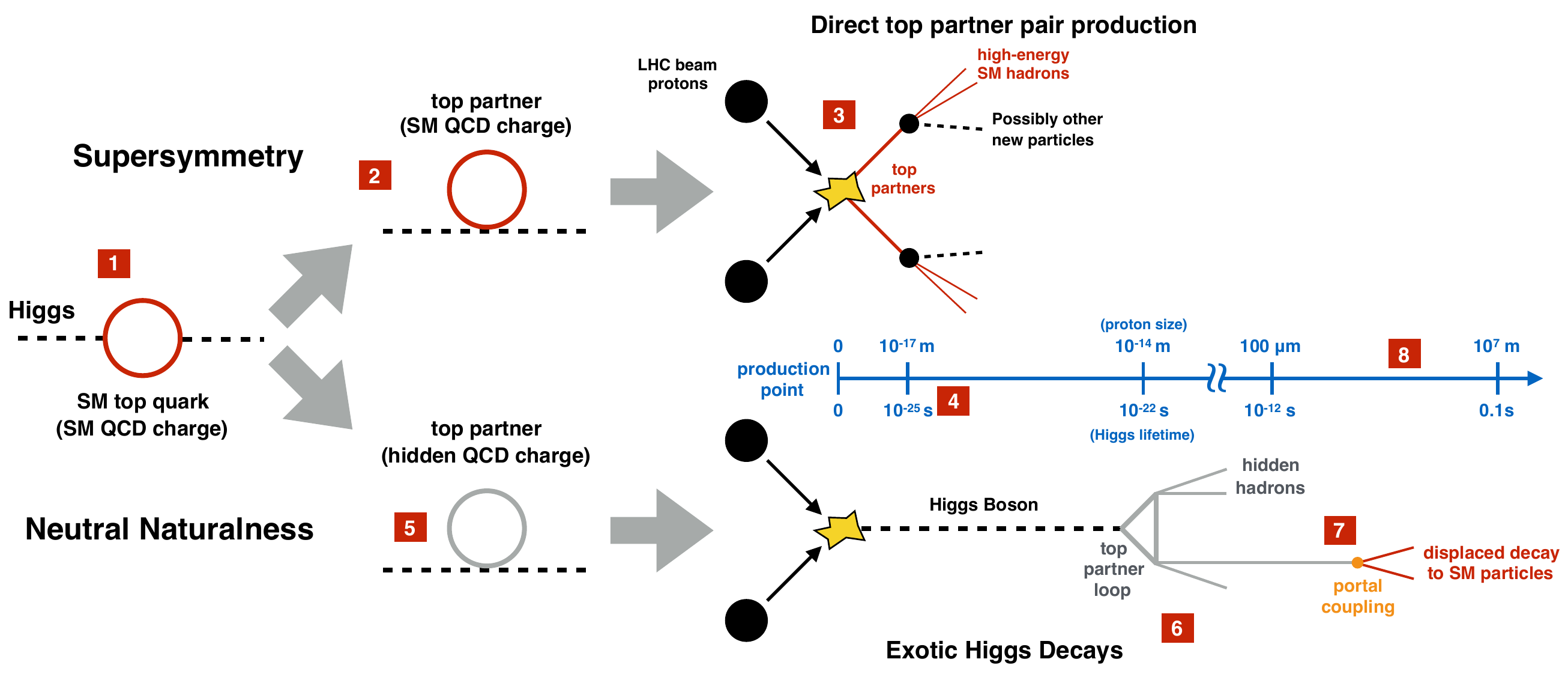}
\end{center}
\justify
\normalsize 
The Hierarchy Problem arises because the Higgs Mass is not protected against large quantum corrections in the SM.
The most important contributions are top quark loops [1], due to their large coupling to the Higgs. These loops are intermittent quantum mechanical fluctuations of the Higgs into top and anti-top pairs.
In Supersymmetry, the top loop is cancelled by a loop of \emph{top partners} called \emph{stops} [2] with mass comparable to the top quark. 
This symmetry assigns the stops the same QCD charge as the tops, predicting their copious production in proton-proton collisions at the LHC [3]. They typically decay into sprays of highly energetic SM particles (and, possibly, new invisible particles) within a very short amount of time [4].
\vspace{3mm}

\noindent Neutral Naturalness\cite{Chacko:2005pe, Craig:2015pha} solves the Hierarchy Problem without SM QCD-charged top partners.
Under a \emph{discrete ``mirror'' symmetry}, the top quark charged under SM QCD is reflected by a top partner charged under a hidden QCD [5].
Due to the absence of SM QCD couplings, top partners would not be produced in large numbers. 
However, due to its coupling to the top partners, the Higgs boson acquires a new decay mode to the hidden QCD hadrons [6]. Eventually these hidden hadrons can decay back to the SM via the Higgs portal [7], showing up as macroscopically [8] displaced decays.
\end{minipage}
}
\end{center}
\caption{ 
\textbf{Neutral Naturalness vs Supersymmetry}
}
\label{b.neutralnaturalness}
\end{table*}

\section*{Solving Mysteries}

\begin{table*}
\begin{center}
\hspace*{-5mm}
\fbox{\begin{minipage}{1.025\textwidth}
\begin{center}
\includegraphics[width=0.78 \textwidth]{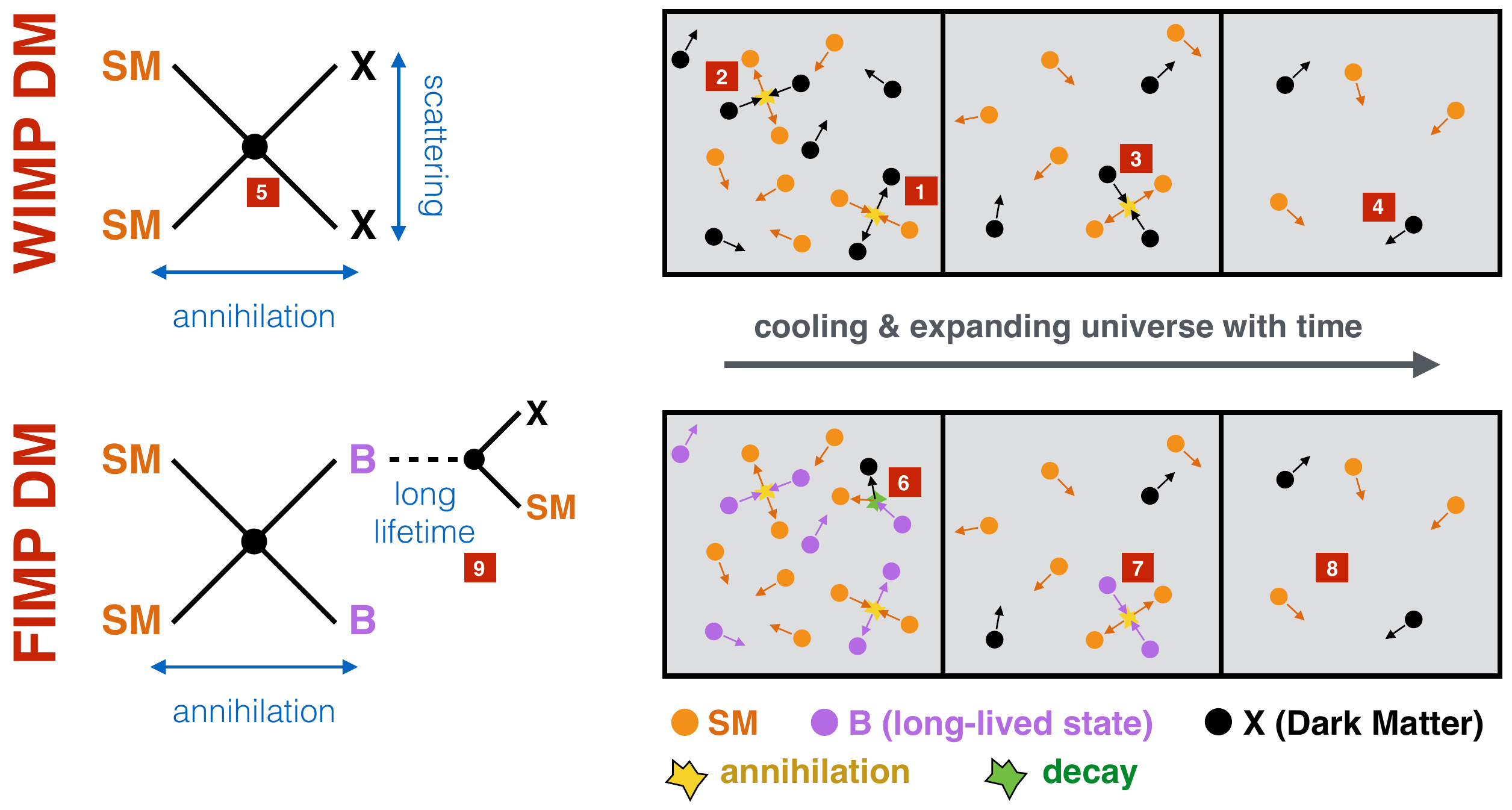}
\end{center}
\justify
\normalsize 

WIMP Dark Matter (top) is a \emph{standard thermal relic}. In the early universe, when the temperature $k T \gg m_X c^2$, both the SM and DM were in thermal equilibrium, with $\mathrm{SM} \ + \ \mathrm{SM} \to X + X$ [1] and the reverse process [2] occurring at equal rates. As the universe cools to $k T \lesssim m_X c^2$, colliding SM particles no longer have the energy to create the heavy DM, but DM still annihilates to the SM [3], drastically reducing its abundance. As the universe cools and expands even more, DM becomes so dilute that annihilations to the SM no longer take place: the abundance is \emph{frozen-out} and survives to this day [4]. This \emph{relic abundance} is dictated by the DM coupling to the SM [5], which sets both the annihilation rate in the early universe, as well as the rate for DM to scatter off SM particles in direct detection experiments. 
\vspace*{3mm}

\noindent The relic density of FIMP Dark Matter (bottom) is set by the \emph{freeze-in} mechanism.\cite{Hall:2009bx} A parent particle $B$ has sizable couplings to the SM and is in thermal equilibrium in the early universe. However, $B$ decays $B \to \mathrm{SM} + X $ with a long lifetime. At this time [6], the universe is much younger than that lifetime, but a tiny fraction of $B$'s still decay to DM, allowing it to steadily accumulate. As the universe cools, $B$ annihilates almost completely away to the SM [7], leaving the frozen-in DM abundance to survive to the present day [8]. In this scenario, the DM relic density is set directly by the lifetime of $B$ [9], typically in the millisecond ballpark.
\end{minipage}
}
\end{center}
\caption{ 
\textbf{Dark Matter Freeze-Out vs Freeze-In}
}
\label{b.darkmatter}
\end{table*}

\begin{table*}
\begin{center}
\hspace*{-5mm}
\fbox{\begin{minipage}{1.025 \textwidth}
\begin{center}
\includegraphics[width=0.78 \textwidth]{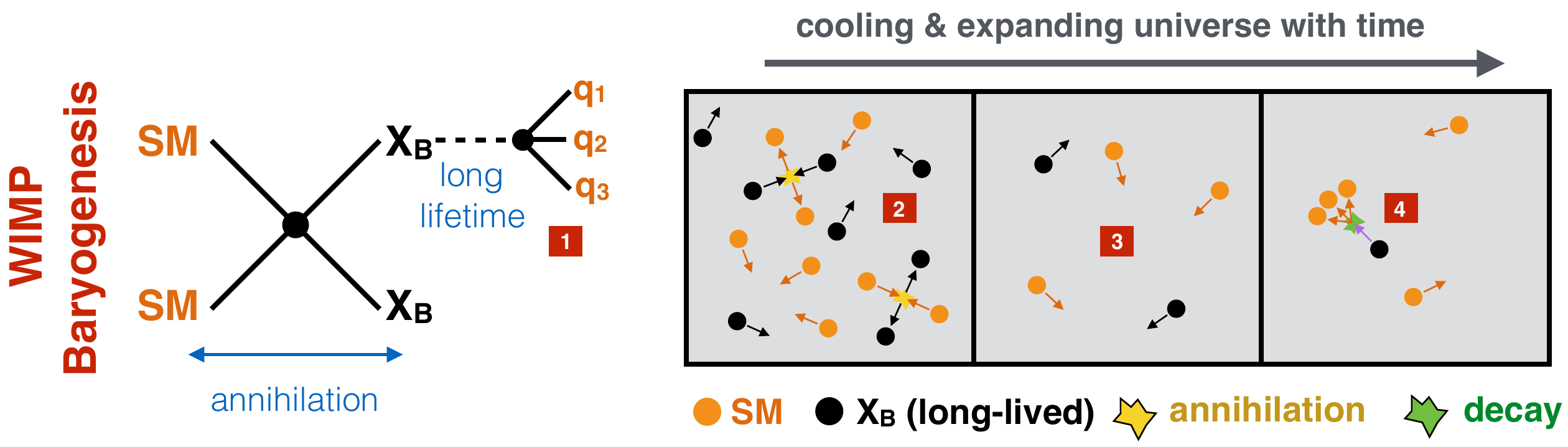}
\end{center}
\justify
\normalsize 
The meta-stable WIMP $X_B$ decays to SM quarks $q_1 q_2 q_3$ [1],  favoring the creation of matter compared to anti-matter.~\cite{Cui:2012jh} Freeze-out of $X_B$ occurs just like for WIMP DM (see \tref{darkmatter}): at early times, $X_B$ is in thermal equilibrium with the matter-antimatter-symmetric SM thermal bath [2] and eventually freezes out as the universe expands and cools [3]. When the universe is about as old as its lifetime, $X_B$ decays {\emph{out-of-equilibrium}} and creates the matter excess [4] which survives to this day. 
\end{minipage}
}
\end{center}
\caption{ 
\textbf{WIMP Baryogenesis}
}
\label{b.baryogenesis}
\end{table*}

Searching for the flashes of LLP decays at the LHC and other colliders will be a major enterprise, requiring dedicated analyses and maybe new detectors, but it is well worth the effort. 
Here are just a a few examples of how new sectors with hidden states can address the three big mysteries.

{Let's start with the Hierarchy Problem. As shown in \tref{neutralnaturalness}, known solutions introduce \emph{top partners} related to the top quark by a symmetry to cancel its large quantum-contribution to the Higgs mass. In canonical solutions like Supersymmetry, the top partners are charged under the SM strong force, or Quantum Chromodynamics (QCD), giving rise to large production rates at the LHC. 
We haven't found any sign of those partners yet, but another solution, called \emph{Neutral Naturalness}\cite{Chacko:2005pe, Craig:2015pha}, relies on Hidden Valleys\cite{Strassler:2006im}.
Hidden Valleys are a family of hidden sector theories that are essentially cousins of SM QCD. They give rise to low-energy bound states called hidden hadrons, in analogy to SM protons and pions.
The top partner is not charged under SM QCD, but is charged under this hidden copy of QCD!
A striking signature involves exotic Higgs decays to hidden hadrons, see \tref{neutralnaturalness}. These can eventually decay back to SM particles via one of several portals, giving rise to displaced decays. }

What about Dark Matter? Perhaps the most minimal and best-known candidate is the \emph{WIMP} (Weakly Interacting Massive Particle). As illustrated in \tref{darkmatter}, it \emph{freezes out} in the early universe with a \emph{relic abundance} set by its coupling to the SM. The ``WIMP miracle'' refers to the observation that electroweak-strength coupling to the SM, with a DM mass near the electroweak scale, roughly generates the  DM abundance we measure today. However, this direct coupling also predicts signatures at direct and indirect detection experiments, which we are still searching for. {At colliders, DM can be produced but is invisible, possibly showing up as a momentum imbalance in the collision.}

A related example of non-minimal Dark Sectors generating LLP signatures is FIMP DM\cite{Hall:2009bx} (Feebly Interacting Massive Particles), with much weaker coupling to the SM. In this case, as illustrated in \tref{darkmatter}, the relic abundance of DM could be set by the lifetime of a heavier parent particle which produces DM in its decays. This parent LLP can be produced at colliders, and lifetimes in the millisecond ballpark are typical.

Finally, Baryogenesis. It can proceed via several known mechanisms. An out-of-equilibrium process needs to create excess matter compared to antimatter in the plasma of the early universe. A very simple way to achieve this is the \emph{out-of-equilibrium decay} of a heavy particle. In particular, the scenario of \emph{WIMP Baryogenesis}\cite{Cui:2012jh}, see \tref{baryogenesis}, piggybacks off the quantitative success of the WIMP miracle. Since the abundances of DM and SM matter in our universe are only different by a factor of 5, a long-lived WIMP $X_B$ decaying into SM particles could easily generate the required matter asymmetry. $X_B$ can also be produced at colliders, giving rise to LLP signatures.

This list of theories is by no means exhaustive, but it proves the point that long-lived particles and hidden sectors are highly motivated for a variety of  fundamental reasons.

\begin{table*}
\begin{center}
\fbox{\begin{minipage}{\textwidth}
\begin{center}
\begin{tabular}{m{0.67\textwidth}m{0.3 \textwidth}}
\includegraphics[width=0.65 \textwidth]{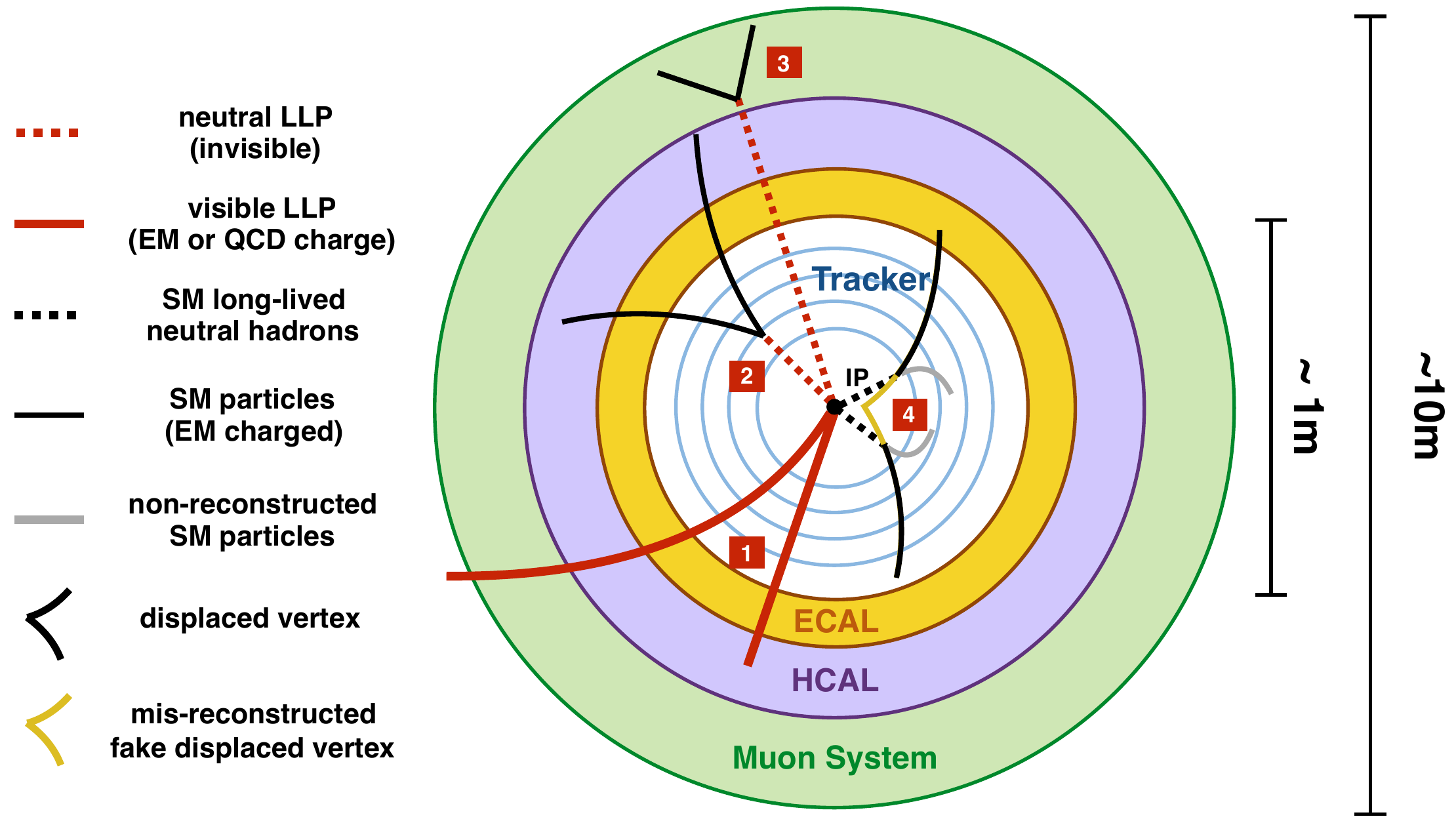}
&
\normalsize
Schematic cross section, transverse to the beam, of the ATLAS and CMS detectors. The inner layers are enlarged to show detail. The tracker layers detect passing charged particles, and track curvature in the magnetic field reveals the particle's momentum. The ECAL and HCAL measure energy of electrically charged particles and hadrons respectively, and the Muon System detects Muons. 

\end{tabular}
\end{center}
\justify
\normalsize 

\noindent Particles are produced in collisions at the interaction point (IP). A few possible LLP signals are shown. Charged LLPs leave strong signals in many detector layers [1]. Neutral LLPs are invisible, but can decay to SM particles in the tracker [2]. The tracks of the decay products reconstruct a \emph{displaced vertex} (DV). At ATLAS, the Muon System can also reconstruct DVs [3]. A rare but important background to neutral LLP searches are \emph{fake displaced vertices}, which occur due to freak QCD events. For example, the tracker might fail to detect all the decay products of long-lived SM hadrons. This is extremely rare, but can lead to a fake DV, shown as yellow intersecting lines [4].
\end{minipage}
}
\end{center}\caption{
\textbf{LLP Searches at the LHC}
}
\label{b.LLPmaindetector}
\end{table*}

\section*{Flashes at Colliders -- why now?}

LLPs can only be reliably identified as such if we know where they were produced. Unlike Dark Matter or cosmic rays, which occur naturally, LLPs have to be synthesized at colliders in order to measure the macroscopic distance from the production point to the flash point of the decay. 

If the LLPs have masses below the electroweak scale, they might be produced in high \emph{intensity} experiments with lower collision energy than the LHC. Examples include the BaBar detector\cite{Lees:2014xha} at the PEP-II $e^+ e^-$ collider, or the APEX\cite{Abrahamyan:2011gv} and SHiP\cite{Alekhin:2015byh} beam dumps. The abundance of collisions allows tiny couplings to be probed by sheer force of numbers. 

The LHC, however, is unique in that it allows us to probe Hidden Sectors with $m_\mathrm{mediator}$ or $m_\mathrm{hidden}$ at or above the electroweak scale. 
The High-Luminosity HL-LHC upgrade, planned for the mid-2020s, will increase the number of collisions by another factor of 10, yielding $10^8$ produced Higgs Bosons. 
A tiny Higgs portal can {easily} force 0.1\% of them to decay to hidden sector states, producing plenty of LLPs. 
Just as  the enormous neutrino fluxes at nuclear reactors paved the way to their detection in the 1950s, the combination of high energy and high intensity collisions in the current era may give us unparalleled access to new hidden worlds, provided we look in the right places.

The LHC main detectors are a busy place, with lots of hadronic shrapnel flying around. Luckily, neutral LLP decays are a spectacular signature, and the burst of energy appearing out of nowhere sets it apart from the mundane rubble emanating from the collision point. 
Looking for LLPs represents something of a paradigm shift from the usual approach to hunting new physics, exemplified by QCD-charged top-partner signatures in \tref{neutralnaturalness}. 
Heavy new physics, with large coupling to the SM, has large production rates, \emph{and they are needed} to observe their signals above the SM background noise, because the same large coupling also makes them decay immediately at the collision point. 
LLP production typically occurs at much lower rates, but each individual displaced decay is so spectacular that backgrounds are orders of magnitude lower. Far fewer observed events are needed to claim discovery.

A priori, the decay lengths of Hidden Sector LLPs could be pretty much anything, from the nearly-prompt (a few hundred microns) to astrophysical length scales. You might think this is a little discouraging -- why should we expect to see an LLP decay in the lab if it could just as well decay near Proxima Centauri? While there can be more detailed theoretical motivations for shorter lifetimes, there is a robust general ceiling. 
If the LLP is producible at colliders, it must have been in thermal equilibrium with the SM plasma in the early universe. As the universe cooled, the first elements -- mostly Hydrogen, Helium, Lithium and Beryllium -- were formed. This process of \emph{Big Bang Nucleosynthesis} (BBN) took place when the universe was about 0.1 - 1 seconds old, and is exquisitely sensitive to the ambient conditions. With few exceptions, LLPs decaying during or after BBN  would disrupt the process\cite{Jedamzik:2006xz}, which is ruled out by measurements of primordial elemental abundances. 
This essentially makes the LLP parameter space finite. Can we probe up to this ceiling with our experiments? We might.

\begin{table*}
\begin{center}
\fbox{\begin{minipage}{\textwidth}
\hspace*{-3mm}
\begin{tabular}{m{0.34\textwidth}cm{0.6\textwidth}}
\normalsize 
\justify
The MATHUSLA\cite{Chou:2016lxi} concept envisions a 200$\times$200 meter detector on the surface above the LHC, slightly offset from the ATLAS or CMS collision points, with a height of roughly 20 m. An LLP produced at the LHC travels upwards and decays in the air-filled decay volume [1]. The detailed inset [2] shows a simulated LLP with mass $20 \gev/c^2$ decaying into a quark and an antiquark, giving rise to an upward traveling shower of $\sim 10$ charged SM particles which are detected by trackers in the roof to reconstruct a DV. The spectacular nature of this signal is vital in rejecting backgrounds.
&
\phantom{bla}
&
\includegraphics[width=0.6 \textwidth]{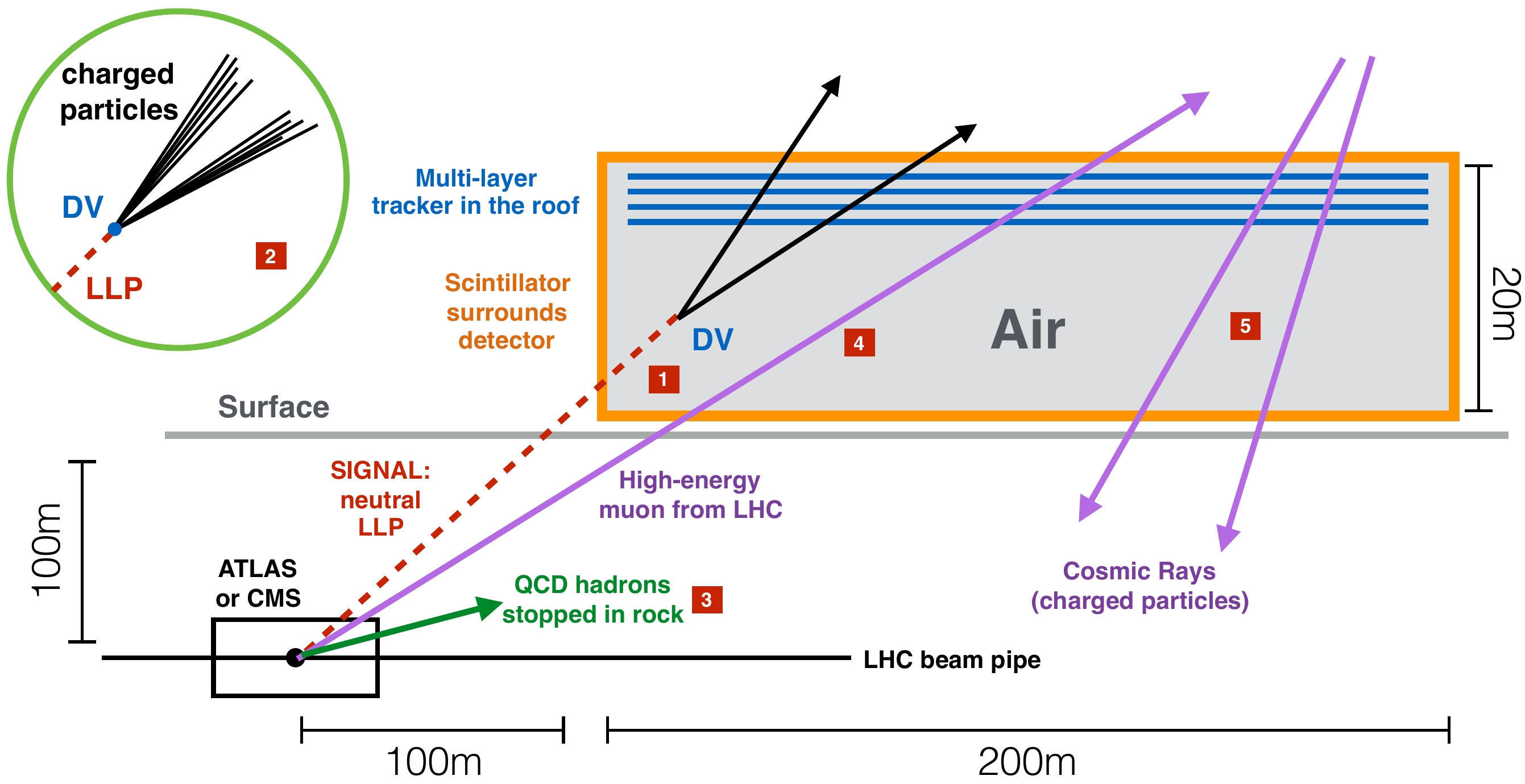}
\end{tabular}
\justify
\normalsize 
\noindent 
SM Hadrons produced in the collision are stopped in the rock [3].  A scintillator veto surrounding the detector volume ensures the DV does not come from upwards traveling charged muons penetrating the rock [4]. Cosmic rays [5] are constantly incident on the whole detector, but they do not reconstruct a DV, and can be {reliably} distinguished by their direction of travel.
Rare backgrounds like cosmic ray neutrinos scattering off air (not shown) 
can be rejected due to the geometry and timing of their final states, which are quite different from the signal shown in [2].
\end{minipage}
}
\end{center}\caption{
\textbf{The MATHUSLA Detector for Ultra-Long Lived Particles}
}
\label{b.mathusla}
\end{table*}

\section*{Searches at the LHC}

Some LLP searches are already being conducted at the LHC, {but for a long time, these signatures were relatively overlooked. Much more still} needs to be done to cover the whole range of accessible masses, lifetimes and production processes. \tref{LLPmaindetector} provides a very basic overview. 

{
LLPs which carry electric or QCD charge are motivated by many theories (with or without hidden sectors), and show up in several   detector subsystems\cite{ATLAS:2014fka, Khachatryan:2016sfv}. 
For example, the LLP could deposit a lot of energy in the calorimeters due to its high mass, or leave a ``kinked track'' if it decays while it traverses the detector. 
}

{
It is also possible for stable hidden sector particles to be electrically milli-charged. This occurs when these states interact with \emph{each other} via a massless hidden photon, which oscillates into our SM sector via the photon portal. These milli-charged states are extremely difficult to detect, since they are only very weakly ionizing when passing through matter. However, their traces could be observed with a dedicated detector like MilliQan\cite{Ball:2016zrp}, placed in the tunnels close to the main detectors.
}

Neutral LLPs are very challenging to isolate, since the detectors weren't primarily designed for them, but if they decay to charged particles, the tracking systems can reconstruct \emph{displaced vertices}\cite{CMS:2014wda} (DV).  These signals are spectacular, but there is \emph{some} SM background. Ultra-rare QCD events, like peculiar configurations of long-lived SM hadron decays and detector hiccups, can line up to fake a DV. The only reason these chains of coincidences can be a problem is because the rate of SM QCD events, which produce multitudes of hadrons at the LHC, is so enormous. 

The obvious challenge are neutral LLPs with decay lengths $c \tau$ much larger than the detector size $L$, in which case only a small fraction $\sim L/(c \tau)$ decay in the detector. 
ATLAS is able to look for DVs in its largest sub-detector, the Muon System. If we produce enough of them we can afford for the large majority to escape, so these searches can have sensitivity to lifetimes in the kilometer range or so\cite{Aad:2015uaa, Coccaro:2016lnz}, but for longer lifetimes (and hence lower chances of decaying in the main detector), QCD background swamps the signal. How can we probe ultra-long lived particles? Is there a way to scratch the BBN ceiling of lifetime parameter space?

\section*{The Lifetime Frontier}

The BBN limit of $\tau \lesssim 0.1 - 1$~s on lifetime allows a produced LLP to travel all the way to the moon. We won't catch them all. Luckily we don't have to. Being able to observe just one in a million would already allow us to probe the most important New Physics scenarios. The main LHC detectors can't do it\ldots What about a dedicated LLP detector?

{
A preliminary concept for a dedicated LLP detector has recently been proposed by one of the authors, together with collaborators. \mbox{\emph{MATHUSLA}}\cite{Chou:2016lxi} (Massive Timing Hodoscope for Ultra-Stable neutraL pArticles), in reference to the long-lived biblical character, is shown in \tref{mathusla}.
Situated on the surface, it is shielded from troublesome SM hadrons produced in the LHC main collision. This gigantic tracker can reconstruct LLP decays   while backgrounds, which are currently being studied in detail, can be near-perfectly rejected. Since the instrumentation is relatively simple, MATHUSLA could be built in time for the High Luminosity LHC upgrade!
}

How useful would this be for discovering new physics? Well, many hidden sectors could go undiscovered at the LHC without a dedicated LLP detector, and the achievable sensitivity of MATHUSLA is remarkable.  For example, if the Higgs decays to LLPs 10\% of the time, the HL-LHC main experiments could detect this as a deviation from SM predictions, but we would not know if we just discovered Dark Matter or LLPs.  MATHUSLA could catch such LLPs decaying, even if their lifetime is right near the BBN limit! For shorter lifetimes (down to $c\tau \sim 100$ meters) the sensitivity is even better, allowing the detection of LLP production rates \emph{three orders of magnitude} smaller than what ATLAS or CMS could discover on their own.


We're in for an exciting ride. 
The LHC is the experiment of our time, 
and there are strong theoretical and empirical clues for new physics near the TeV scale. 
Null results of searches to date don't invalidate those motivations. 
On the contrary, they may just be pointing us towards the brave new world of Hidden Sectors. 
Much work is needed to take advantage of the unique opportunities the LHC provides, but if the necessary searches and detectors are implemented now, 
we have every reason to hope for revolutionary discoveries in the years to come.

\bibliographystyle{unsrt}
\bibliography{darksectors}

\begin{thebibliography}{10}

\bibitem{Bjorken:2009mm}
James~D. Bjorken, Rouven Essig, Philip Schuster, and Natalia Toro.
\newblock {New Fixed-Target Experiments to Search for Dark Gauge Forces}.
\newblock {\em Phys. Rev.}, D80:075018, 2009.

\bibitem{Strassler:2006ri}
Matthew~J. Strassler and Kathryn~M. Zurek.
\newblock {Discovering the Higgs through highly-displaced vertices}.
\newblock {\em Phys. Lett.}, B661:263--267, 2008.

\bibitem{Curtin:2013fra}
David Curtin et~al.
\newblock {Exotic decays of the 125 GeV Higgs boson}.
\newblock {\em Phys. Rev.}, D90(7):075004, 2014.

\bibitem{Chacko:2005pe}
Z.~Chacko, Hock-Seng Goh, and Roni Harnik.
\newblock {The Twin Higgs: Natural electroweak breaking from mirror symmetry}.
\newblock {\em Phys. Rev. Lett.}, 96:231802, 2006.

\bibitem{Craig:2015pha}
Nathaniel Craig, Andrey Katz, Matt Strassler, and Raman Sundrum.
\newblock {Naturalness in the Dark at the LHC}.
\newblock {\em JHEP}, 07:105, 2015.

\bibitem{Hall:2009bx}
Lawrence~J. Hall, Karsten Jedamzik, John March-Russell, and Stephen~M. West.
\newblock {Freeze-In Production of FIMP Dark Matter}.
\newblock {\em JHEP}, 03:080, 2010.

\bibitem{Cui:2012jh}
Yanou Cui and Raman Sundrum.
\newblock {Baryogenesis for weakly interacting massive particles}.
\newblock {\em Phys. Rev.}, D87(11):116013, 2013.

\bibitem{Strassler:2006im}
Matthew~J. Strassler and Kathryn~M. Zurek.
\newblock {Echoes of a hidden valley at hadron colliders}.
\newblock {\em Phys. Lett.}, B651:374--379, 2007.

\bibitem{Lees:2014xha}
J.~P. Lees et~al.
\newblock {Search for a Dark Photon in $e^+e^-$ Collisions at BaBar}.
\newblock {\em Phys. Rev. Lett.}, 113(20):201801, 2014.

\bibitem{Abrahamyan:2011gv}
S.~Abrahamyan et~al.
\newblock {Search for a New Gauge Boson in Electron-Nucleus Fixed-Target
  Scattering by the APEX Experiment}.
\newblock {\em Phys. Rev. Lett.}, 107:191804, 2011.

\bibitem{Alekhin:2015byh}
Sergey Alekhin et~al.
\newblock {A facility to Search for Hidden Particles at the CERN SPS: the SHiP
  physics case}.
\newblock {\em Rept. Prog. Phys.}, 79(12):124201, 2016.

\bibitem{Jedamzik:2006xz}
Karsten Jedamzik.
\newblock {Big bang nucleosynthesis constraints on hadronically and
  electromagnetically decaying relic neutral particles}.
\newblock {\em Phys. Rev.}, D74:103509, 2006.

\bibitem{Chou:2016lxi}
John~Paul Chou, David Curtin, and H.~J. Lubatti.
\newblock {New Detectors to Explore the Lifetime Frontier}.
\newblock {\em Phys. Lett.}, B767:29--36, 2017.

\bibitem{ATLAS:2014fka}
Georges Aad et~al.
\newblock {Searches for heavy long-lived charged particles with the ATLAS
  detector in proton-proton collisions at $ \sqrt{s}=8 $ TeV}.
\newblock {\em JHEP}, 01:068, 2015.

\bibitem{Khachatryan:2016sfv}
Vardan Khachatryan et~al.
\newblock {Search for long-lived charged particles in proton-proton collisions
  at sqrt(s) = 13 TeV}.
\newblock 2016.

\bibitem{Ball:2016zrp}
Austin Ball et~al.
\newblock {A Letter of Intent to Install a milli-charged Particle Detector at
  LHC P5}.
\newblock 2016.

\bibitem{CMS:2014wda}
Vardan Khachatryan et~al.
\newblock {Search for Long-Lived Neutral Particles Decaying to Quark-Antiquark
  Pairs in Proton-Proton Collisions at $\sqrt{s} =$ 8 TeV}.
\newblock {\em Phys. Rev.}, D91(1):012007, 2015.

\bibitem{Aad:2015uaa}
Georges Aad et~al.
\newblock {Search for long-lived, weakly interacting particles that decay to
  displaced hadronic jets in proton-proton collisions at $\sqrt{s}=8$ TeV with
  the ATLAS detector}.
\newblock {\em Phys. Rev.}, D92(1):012010, 2015.

\bibitem{Coccaro:2016lnz}
Andrea Coccaro, David Curtin, H.~J. Lubatti, Heather Russell, and Jessie
  Shelton.
\newblock {Data-driven Model-independent Searches for Long-lived Particles at
  the LHC}.
\newblock {\em Phys. Rev.}, D94(11):113003, 2016.

\end{thebibliography}

\end{document}